\documentstyle[12pt]{article}

\catcode`\@=11 \long\def\@makefntext#1{ \protect\noindent \hbox to
3.2pt {\hskip-.9pt
$^{{\ninerm\@thefnmark}}$\hfil}#1\hfill}                

\def\@makefnmark{\hbox to 0pt{$^{\@thefnmark}$\hss}}  

\def\ps@myheadings{\let\@mkboth\@gobbletwo
\def\@oddhead{\hbox{}
\rightmark\hfil\ninerm\thepage}
\def\@oddfoot{}\def\@evenhead{\ninerm\thepage\hfil
\leftmark\hbox{}}\def\@evenfoot{}
\def\sectionmark##1{}\def\subsectionmark##1{}}

\renewcommand{\thefootnote}{\fnsymbol{footnote}}

\newcounter{sectionc}\newcounter{subsectionc}\newcounter{subsubsectionc}
\renewcommand{\section}[1] {\vspace*{0.6cm}\addtocounter{sectionc}{1}
\setcounter{subsectionc}{0}\setcounter{subsubsectionc}{0}\noindent
        {\normalsize\bf\thesectionc. #1}\par\vspace*{0.4cm}}
\renewcommand{\subsection}[1] {\vspace*{0.6cm}\addtocounter{subsectionc}{1}
        \setcounter{subsubsectionc}{0}\noindent
        {\normalsize\it\thesectionc.\thesubsectionc. #1}\par\vspace*{0.4cm}}
\renewcommand{\subsubsection}[1]
{\vspace*{0.6cm}\addtocounter{subsubsectionc}{1}
        \noindent
{\normalsize\rm\thesectionc.\thesubsectionc.\thesubsubsectionc.
        #1}\par\vspace*{0.4cm}}

\renewenvironment{thebibliography}[1]
        {\begin{list}{\arabic{enumi}.}
        {\usecounter{enumi}\setlength{\parsep}{0pt}
\setlength{\leftmargin 0.52cm}{\rightmargin 0pt}
         \setlength{\itemsep}{0pt} \settowidth
        {\labelwidth}{#1.}\sloppy}}{\end{list}}

\newcounter{itemlistc}
\newcounter{romanlistc}
\newcounter{alphlistc}
\newcounter{arabiclistc}

\def\@citex[#1]#2{\if@filesw\immediate\write\@auxout
        {\string\citation{#2}}\fi
\def\@citea{}\@cite{\@for\@citeb:=#2\do
        {\@citea\def\@citea{,}\@ifundefined
        {b@\@citeb}{{\bf ?}\@warning
        {Citation `\@citeb' on page \thepage \space undefined}}
        {\csname b@\@citeb\endcsname}}}{#1}}

\newif\if@cghi
\def\cite{\@cghitrue\@ifnextchar [{\@tempswatrue
        \@citex}{\@tempswafalse\@citex[]}}
\def\citelow{\@cghifalse\@ifnextchar [{\@tempswatrue
        \@citex}{\@tempswafalse\@citex[]}}
\def\@cite#1#2{{$\null^{#1}$\if@tempswa\typeout
        {IJCGA warning: optional citation argument
        ignored: `#2'} \fi}}

 1 
scaled\magstep 1  1
 
scaled\magstephalf

 \font\ninerm=cmr9

\newcommand{\beq}{\begin{equation}}
\newcommand{\eeq}{\end{equation}}

\newcommand\mathC{\mkern1mu\raise2.2pt\hbox{$\scriptscriptstyle|$}
                {\mkern-7mu\rm C}}

\begin{document}

\hfill \vspace*{1cm}

\centerline{\normalsize\bf MAXIMUM ENTROPY MULTIVARIATE DENSITY
ESTIMATION:} \centerline{\normalsize\bf AN EXACT GOODNESS-OF-FIT
APPROACH}

\vspace*{0.6cm} \centerline{\footnotesize
SABBIR~RAHMAN\footnote{E-mail: sarahman@alum.mit.edu} ~\&
MAHBUB~MAJUMDAR\footnote{E-mail: m.majumdar@imperial.ac.uk}}
\vspace*{0.2cm} \baselineskip=13pt \centerline{\footnotesize\em
Theoretical Physics, Blackett Laboratory}
\centerline{\footnotesize\em Imperial College of Science,
Technology \& Medicine} \centerline{\footnotesize\em Prince
Consort Road, London SW7 2BZ, U.K.}\vspace*{0.9cm}

{\centering{\begin{minipage}{12.2truecm}\footnotesize\baselineskip=12pt
\noindent\centerline{\footnotesize ABSTRACT} \vspace*{0.1cm}
\parindent=0pt

We consider the problem of estimating the population probability
distribution given a finite set of multivariate samples, using the
maximum entropy approach. In strict keeping with Jaynes' original
definition, our precise formulation of the problem considers
contributions only from the smoothness of the estimated
distribution (as measured by its entropy) and the loss functional
associated with its goodness-of-fit to the sample data, and in
particular does not make use of any additional constraints that
cannot be justified from the sample data alone. By mapping the
general multivariate problem to a tractable univariate one, we are
able to write down exact expressions for the goodness-of-fit of an
arbitrary multivariate distribution to any given set of samples
using both the traditional likelihood-based approach and a
rigorous information-theoretic approach, thus solving a
long-standing problem. As a corollary we also give an exact
solution to the `forward problem' of determining the expected
distributions of samples taken from a population with known
probability distribution.

\end{minipage}\par}}
\vspace*{0.6cm}

\normalsize\baselineskip=15pt \setcounter{footnote}{0}
\renewcommand{\thefootnote}{\alph{footnote}}    

\section{Introduction}

According to Jaynes\cite{Jaynes57}, the maximum entropy
distribution is ``uniquely determined as the one which is
maximally noncommittal with regard to missing information, in that
it agrees with what is known, but expresses maximum uncertainty
with respect to all other matters"\cite{Wu03}.

On the other hand, Kapur and Kesavan\cite{Kapur92} state that
``the maximum entropy distribution is the most unbiased
distribution that agrees with given moment constraints because any
deviation from maximum entropy will imply a bias".

While the latter neatly encapsulates the modern interpretation of
the maximum entropy principle in its application to density
estimation, it is not equivalent to the definition given by Jaynes
as it restricts its use to the case where the moments of the
population distribution are already known.

While this restriction may be convenient, it is not valid in any
case in which one is not simply trying to re-derive a standard
distribution based upon its known moments using maximum entropy
principles. Rather, in practical applications the moments of the
population distribution are not (and indeed cannot) be known a
priori, and certainly cannot be determined on the basis of a
finite number of samples.

In this paper, we give an explicit and exact expression of the
maximum entropy density estimation problem in a form which is
strictly in keeping with Jaynes' original (and precise)
definition.

\section{Reformulating the MaxEnt Problem}

So let us return to basics and consider the problem of estimating
the multivariate population density distribution given a finite
set of samples taken at random from the population, assuming that
the raw sample data is the {\it only} prior information we have.
In this case, which is clearly of the most general practical
applicability, the requirement that the maximum entropy
distribution `agrees with what is known' is equivalent to the
requirement that the population distribution provides a good fit
to the sample data. In this sense the maximum entropy distribution
can be defined as ``the distribution of maximum entropy subject to
the provision of a good fit to the sample data", with the only
potential uncertainty lying in the relative importance which
should be attached to each of the two contributions. While this
uncertainty reflects the supposed ill-posedness of the density
estimation problem, Jaynes' definition implies that there should
in fact exist a unique solution, so that even this uncertainty is
in principle resolvable. While we do attempt to resolve this issue
here, the matter certainly deserves further attention.

The definition given in the last paragraph allows us to formulate
the maximum entropy multivariate density estimation problem in
precise mathematical terms. If we denote the estimated
distribution by $f(r)$ where $r\in R^D$, and the sample data set
by $\{x_1,\dots,x_N\}$, we would like to maximise the functional
defined by,

\begin{equation}
F[f(r)] = S[f(r)] + \alpha G[f(r),\{x_i\}]\,,\label{eqn:maxprob}
\end{equation}
where $S[f(r)]$ is the normalised\footnote{The normalisation
factor is fixed by requiring that the entropy of the uniform
distribution be unity.} , sample-independent entropy of the
estimated distribution over its domain of definition,

\begin{equation}
S[f(r)]=-{1\over\log V}\int f(r)\log f(r)\,
d\tau\,,\qquad(\hbox{where } V=\int d\tau)\,,\label{eqn:norment}
\end{equation}
and $G[f(r),\{x_i\}]$ is a precise measure of the goodness-of-fit
of the distribution to the sample data. An optional tunable
variable $\alpha\in[0,\infty]$ has been included which
parametrises the solutions. It is clear by inspection that
$\alpha=0$ implies the sample-independent maximum entropy solution
represented by the uniform distribution $f(r)=$ constant, while
the limit $\alpha\rightarrow\infty$ corresponds to the
distribution providing the best fit to the data without regard to
its entropy. The solution for any other value of $\alpha$ will
represent some trade-off between maximising entropy and maximising
the goodness-of-fit. The fact that neither of the two extremal
solutions would be of use in practical applications does support
the argument that there should exist an optimal value for $\alpha$
(presumably unity), and hence a unique optimal density estimate.
We will come back to this point later.

\section{Establishing the Goodness-of-Fit}

We have yet to give the expression for the goodness-of-fit
$G[f(r),\{x_i\}]$. In the absence of an analytically rigorous and
generally applicable measure of goodness-of-fit, various ad hoc
schemes have been used in the past\cite{Agostino86,Aslan02}. As we
will show, there do exist unique analytical expressions for the
goodness-of-fit of an arbitrary multivariate probability
distribution $f(r)$ to a given set of sample data
$\{x_1,\dots,x_N\}$ depending on whether a likelihood-based or
information-theoretic approach is used. While the former does
correctly provide the likelihood of obtaining any particular set
of sample values assuming a given population distribution, we will
nevertheless demonstrate that it is the information-theoretic
approach that is the appropriate one to use to find the population
distribution which best accounts for the samples observed.

\subsection{Mapping Multivariate Estimation to a Univariate Problem}

It happens that there exists a well-defined procedure for mapping
the complex multivariate problem into a tractable univariate one.
To proceed, one needs to note that the probability of a sample
taking values in a particular region of $R^D$ is given by the area
(or more generally the hypervolume) under the curve $f(r)$ over
that region. Moreover we know that for a probability distribution,
the total area under the curve is normalised to unity.

The key step is to define a mapping $C_f:R^D\rightarrow I$
(representing a particular kind of cumulative probability density
function corresponding to $f(r)$) from $R^D$ onto the real line
segment $I=[0,1]$ as follows,

\begin{equation}
C_f(x) = \int f(r) \Theta[f(x)-f(r)] d\tau\,,\label{eqn:cumpdf}
\end{equation}
where $x\in R^D$ and $\Theta(y)$ is the Heaviside step function
with $\Theta(y)=1$ for $y\geq0$ and $\Theta(y)=0$ otherwise. The
mapping $C_f$ will in general be many-to-one. Its utility lies in
the fact that if we take the set of samples $\{x_1,\dots,x_N\}$ in
$R^D$ and map them to the the set of points
$\{C_f(x_1),\dots,C_f(x_N)\}$ on the segment $I$ then, in view of
the equivalence between the probability and the area under the
curve, the goodness-of-fit of $f(r)$ to the samples $\{x_i\}$ is
precisely equal to the goodness-of-fit of the uniform probability
distribution $g(x)=1$ defined on the segment $I$ to the mapped
samples $\{C_f(x_i)\}$. Let us now consider the latter case in
more detail.

\subsection{Uniformly Distributed Samples on a Real Line Segment}

Consider a perfect random number generator which generates values
uniformly distributed in the range $[0,1]$. Suppose we plan to use
it to generate $N$ random samples. We can calculate in advance the
probability distribution $p_{N,i}(x)$ of the $i$-th sample (where
the samples are labelled in order of increasing magnitude), as
follows.

Let $X_i$ be the random variable corresponding to the value of the
$i$-th sample for each $i=1\dots N$. Note that the probability of
a number (selected at random from $[0,1]$ assuming a uniform
distribution) being less than some value $x\in[0,1]$ is simply
$x$, while the probability of it being greater than $x$ is $1-x$.
Thus, if we consider the $i$-th value in a set of $N$ samples
taken at random, the probability that $X_i$ takes the value $x$ is
given by the product of the probability $x^{i-1}$ that $i-1$ of
the values are less than $x$ and the probability $(1-x)^{N-i}$
that the remaining $N-i$ values are greater than $x$, divided by a
combinatorial factor $Z_{N,i}$ counting the number of ways $N$
integers can be partitioned into three sets of size $i-1$, $1$ and
$N-i$ respectively,

\begin{equation}
p_{N,i}(x) \equiv P(X_i=x) = Z_{N,i}^{-1} \, x^{i-1}
(1-x)^{N-i}\,.\label{eqn:pdfs}
\end{equation}
From simple combinatorics, the value of $Z_{N,i}$ is given by,

\begin{equation}
Z_{N,i} = {N!\over(i-1)!(N-i)!} =
{\Gamma(N+1)\over\Gamma(i)\Gamma(N-i+1)} =
B^{-1}(N-i+1,i)\,,\label{eqn:betafn}
\end{equation}
where $B(p,q)$ is the Euler beta function which appears in the
Veneziano amplitude for string scattering\cite{Green87}. That this
value is correct can be checked using the fact that $p_{N,i}(x)$
must be normalised so that $\int p_{N,i}(x) dx = 1$, and noting
that the resulting integral is just the definition of the beta
function given above. Note also that if experiments are carried
out in which sets of $N$ samples are taken repeatedly, the
expectation of the $i$-th sample is given by,

\begin{equation}
E[X_i] = \int_0^1 x p_{N,i}(x)\, dx = {i\over
N+1}\,,\label{eqn:sampdist}
\end{equation}
for $i=1\dots N$, corresponding to the most regularly distributed
configuration of the $N$ samples possible, and in excellent accord
with intuition.

\subsection{The Maximum Likelihood Approach}

Taking a traditional likelihood-based approach, an expression for
the goodness-of-fit of a set of $N$ samples to the uniform
distribution on $[0,1]$ can now be obtained by first labelling the
samples in order of increasing magnitude and then calculating the
likelihood given by,

\begin{equation}
L[\{x_i\}] = \prod_{i=1}^N p_{N,i}(x_i)\,.\label{eqn:likelihood}
\end{equation}

Bearing in mind the mapping $C_f:R^D\rightarrow I$ defined in
(\ref{eqn:cumpdf}), we can generalise the above to derive an exact
expression for the goodness-of-fit of a set of $N$ samples
$\{x_1,\dots,x_N\}$ to an arbitrary multivariate probability
distribution $f(r)$,

\begin{equation}
L[f(r),\{x_i\}] = L[\{C_f(x_i)\}] = \prod_{i=1}^N
p_{N,i}(C_f(x_i))\,.\label{eqn:gof}
\end{equation}
where the samples are now labelled in order of increasing
magnitude of $f(x_i)$ and hence $C_f(x_i)$.

Let us take a closer look at the likelihood measure of
Eqn.(\ref{eqn:likelihood}) and in particular, let us consider the
simple illustrative case when only two samples are taken from the
uniform distribution on [0,1]. In this case, the likelihood is
maximised if the samples happen to take precisely the values $0$
and $1$. This slightly perturbing result is actually correct and
is one of the reasons why care must be taken if one wishes to
apply likelihood-based arguments in the opposite direction to
estimate the population distribution based upon observed sample
data. More generally, the expression (\ref{eqn:gof}) for the
likelihood will always be biased towards the case where the
position of the first sample coincides with the minimum value of
the probability distribution and that of the last sample with its
maximum.

These considerations are sufficient to show that the maximum
likelihood approach to multivariate density estimation is
problematic and provide us with good reason to seek an
alternative, more rigorous approach.

\subsection{The Information Theoretic Approach}

The rigorous alternative lies in taking an information theoretic
approach. Indeed we will show that it is possible to assign a
unique entropy associated with the goodness-of-fit of the
estimated population distribution to the sample data in the same
way (see Eqn.(\ref{eqn:norment})) that an entropy was assigned to
the estimated distribution itself.

To see how, consider the values $y_i\equiv C_f(x_i)$ of the
samples obtained after having mapped them to the real segment
using the mapping defined in Eqn.(\ref{eqn:cumpdf}). Defining
$y_0=0$ and $y_{N+1}=1$ for convenience, these values are
constrained by the `normalisation' property,

\begin{equation}
\sum_{i=1}^{N+1} y_i - y_{i-1} = 1\,.\label{eqn:datanorm}
\end{equation}

Then by considering each of the $N+1$ gaps between the values as
`sample bins', and the size $d_i\equiv y_i-y_{i-1}$ of each gap as
the probability associated with the corresponding bin, it becomes
possible to identify the distribution of the mapped samples on
$[0,1]$ with a discrete probability distribution defined over the
set of $N+1$ sample bins. The (normalised) entropy associated with
the fit of the estimated population distribution to the sample
data can then be equated with the entropy of the equivalent
discrete probability distribution,

\begin{equation}
S'[f(r),\{x_i\}] \equiv -{1\over\log(N+1)}\sum_{i=1}^{N+1} d_i
\log d_i\,.\label{eqn:dataentropy}
\end{equation}
This is just the discrete version of the expression given in
Eqn.(\ref{eqn:norment}) for a continuous probability distribution.
Maximising Eqn.(\ref{eqn:dataentropy}) for the entropy results
immediately in the desirable property that the samples are equally
spaced, namely $y_i = i/(N+1)$, in agreement with the expected
values obtained less directly in Eqn.(\ref{eqn:sampdist}).

The discussion above strongly suggests that the entropy $S'$ of
Eqn.(\ref{eqn:dataentropy}) should be used instead of the
traditional likelihood (as given here by Eqn.(\ref{eqn:gof})),
both as a measure of the goodness-of-fit of an arbitrary
population distribution to a given set of multivariate samples,
and also as the second term $G[f(r),\{x_i\}]$ appearing in the
functional of Eqn.(\ref{eqn:maxprob}).

Substituting (\ref{eqn:dataentropy}) into (\ref{eqn:maxprob}), we
claim that the rigorous solution to the MaxEnt multivariate
density estimation problem is given by the probability
distribution which maximises the functional,

\begin{equation}
F[f(r)] = -{1\over\log V}\int f(r)\log f(r)\, d\tau -
{\alpha\over\log(N+1)} \sum_{i=1}^{N+1} d_i \log
d_i\,,\label{eqn:final}
\end{equation}
where,
\begin{equation}
d_i = C_f(x_i)-C_f(x_{i-1})\,,\label{eqn:diff}
\end{equation}
and the mapping $C_f$ is given by Eqn.(\ref{eqn:cumpdf}). In
(\ref{eqn:final}) the parameter $\alpha\in[0,\infty]$ can be used
to tune the solutions, bearing in mind that a smaller value will
emphasise the smoothness of the resulting distribution, while a
larger value will emphasise the goodness of fit. Setting $\alpha$
to unity and maximising will in principle give the
unique\footnote{Note that there is some potential for
non-uniqueness to creep in due to the possibility of degenerate
solutions in certain situations such as when $N$ is very small or
when the samples are distributed highly symmetrically.}\, maximum
entropy distribution as originally envisioned by
Jaynes\footnote{The fact that the entropy of a uniform
distribution over $[-\infty,\infty]$ is infinite ($\sim\log V$),
while being finite for other distributions
($\sim\log\sigma\sqrt{2\pi e}$ for a univariate Gaussian),
suggests that special attention may be required in the non-compact
case to prevent one term overwhelming the other. A simple way of
regulating the problem is to consider non-compact domains as
extremal limits of compact ones where the distributions are
constrained to be smooth and to vanish at the boundaries.}\,.

The information-theoretic approach we have described in this
section is more rigorous and compelling than the traditional
likelihood approach, as clearly evidenced by the pleasingly
symmetric form (\ref{eqn:final}) of the resulting optimisation
problem. In particular, both terms contributing to the functional
are associated with entropies - the first term being the entropy
associated with the smoothness of the estimated population
distribution, and the second term being the entropy associated
with the goodness-of-fit of the distribution to the sample data.

Algorithms implementing the optimisation procedure are under
development, which will allow us to calculate specific solutions
of (\ref{eqn:final}) and to perform more detailed investigations
of their properties. We hope to present these results in a future
paper.

\subsection{A Corollary: The Forward Problem}

Before ending, it is worth mentioning here as a corollary that the
distributions $p_{N,i}(x)$ of (\ref{eqn:pdfs}) also help us to
solve the `forward problem', i.e. that of determining the expected
distributions $p^f_{N,i}(r)$ of any set of $N$ samples taken at
random from a multivariate population where the population density
distribution $f(r)$ is given.

\subsubsection{The univariate case}

We would like to know the expected distribution of the samples
when the univariate population distribution $f(x)$ is given. In
place of the mapping of (\ref{eqn:cumpdf}), it is appropriate here
to consider the mapping $C_f'$ defined by the cumulative
probability density function,

\begin{equation}
C_f'(x) = \int_{-\infty}^x f(y)\, dy\,.\label{eqn:newcpdf}
\end{equation}
If the univariate samples are labelled in increasing order of
value then their expected distributions are given by,

\begin{equation}
p^f_{N,i}(x) = p_{N,i}(C_f'(x))\,,\label{eqn:forwarduni}
\end{equation}
and these can be used for example to estimate the experimental
errors in individual sample values given an estimate of the
population distribution\cite{Buck91}.

\subsubsection{The multivariate case}

The forward problem does not have an obvious generalisation to the
multivariate case because of the lack of an unambiguous definition
of the cumulative probability density function in that case.
Nevertheless we can instead apply the mapping $C_f$ of
Eqn.(\ref{eqn:cumpdf}) (paying careful attention to the
degeneracies present) to obtain the following expected
distributions for the samples ordered as described below
Eqn.(\ref{eqn:gof}),

\begin{equation}
p^f_{N,i}(r) = J^{-1}(r)\,p_{N,i}(C_f(r))\,,\label{eqn:forward}
\end{equation}
where $J(r)$ measures the (typically $(D-1)$-dimensional) volume
of the degeneracy of $f(r)$ (i.e. the volume of the subspace of
$R^D$ sharing the same value of $f(r)$) for each value of $r$. At
special values the region of degeneracy may have dimensionality
less than $(D-1)$ in which case the value of $p^f_{N,i}(r)$
becomes irrelevant and can safely be ignored. On the other hand
for distributions which contain $D$-dimensional subspaces
throughout which $f(r)$ is constant (the uniform distribution
being an obvious example), then special considerations will be
required in order to generalise the analysis leading to
Eqn.(\ref{eqn:pdfs}) for the real line segment to irregular,
multidimensional, and possibly non-compact spaces. Excepting the
simplest cases, such an analysis promises to be highly non-trivial
and we will not attempt to delve into such intricacies here. Note
that (\ref{eqn:forward}) does not agree with
(\ref{eqn:forwarduni}) in the univariate case as the labelling of
the samples and the corresponding interpretations of the
distributions are quite different in each case.

It is often assumed that the deviations of individual observations
from their expected values follow a normal
distribution\cite{Buck91} for univariate data, leading to a
$\chi^2$ measure of goodness-of-fit\footnote{A discussion of the
trade-off between smoothness and goodness-of-fit in the context of
this assumption appears in Gull (1989)\cite{Gull89}.}\,. Our exact
results in Eqn.(\ref{eqn:forwarduni}) demonstrate that this
approximation only holds if $N$ is sufficiently large and only
then if $f(r)$ is sufficiently well-behaved. We will conclude our
analysis at this point.

\section{Summary and Discussion}

The purpose of the present work has been to reformulate the
maximum entropy (MaxEnt) density estimation problem in a precise
way which is in strict keeping with its original definition as
introduced by Jaynes. The importance of having such a precise
formulation hardly needs mentioning given the ubiquity of the
estimation problem throughout the sciences.

In reaching our formulation we have managed to solve the
long-standing problem of obtaining an exact expression for the
likelihood of observing any particular set of sample values when
taken at random from a given population. This is useful in the
experimental sciences for validating theoretical models on the
basis of observations. As a corollary, we have also been able to
propose the solution to the `forward problem' - that of
determining the distribution of the samples when the population
distribution is known.

The traditional maximum likelihood approach was shown to have some
unsatisfactory features when applied to the problem of density
estimation. On the other hand, by taking a novel
information-theoretic approach we have succeeded in deriving an
explicit and rigorous entropic measure of the goodness-of-fit of a
generic population distribution to a given set of multivariate
samples. This in turn has made it possible to reformulate the
MaxEnt density estimation problem in a unique, precise and purely
information-theoretic way.

We have made allowance for the introduction of an optional tunable
parameter in our expression of the MaxEnt problem which
parametrises solutions ranging from those with maximal smoothness
to those providing maximal fit to the data. The effect of this
parameter on the solutions has not been discussed in detail here,
and we intend to come back to it in future once computational
algorithms implementing the optimisation have been developed.

\section{Acknowledgements}

We would like to thank Wajid Mannan, Mohamed Mekias and Asif
Khalak for their valuable suggestions during the preparation of
this manuscript.

\end{document}